\documentclass[aps,twocolumn,showpacs,showkeys]{revtex4}
\usepackage{amssymb}
\usepackage{natbib}
\usepackage{amsmath}
\usepackage{amsfonts}
\usepackage{graphicx}
\usepackage{mathrsfs}
\usepackage{dcolumn}
\usepackage{bm}
\usepackage{mathbbol}
\usepackage{color}

\def\Tr{\mbox{Tr}}

\definecolor{rot}{rgb}{0.75,0.05,0.25}
\definecolor{hellgrau}{gray}{0.5}
\definecolor{blau}{rgb}{0,0,0.7}

\begin{document}
\title{Influence of measurements on the statistics of work
performed on a quantum system}
\author{Michele Campisi, Peter
  Talkner, and Peter H\"anggi}
\affiliation{Institute of Physics, University of Augsburg,
Universit\"atsstrasse 1, D-86135 Augsburg, Germany}
\date{\today}

\begin{abstract}
The recently demonstrated robustness of fluctuation theorems
against measurements [M. Campisi \emph{et al.}, Phys. Rev. Lett.
\textbf{105} 140601 (2010)] does not imply that the probability
distributions of nonequilibrium quantities, such as heat and
work, remain unaffected.
We determine the impact of measurements that are performed during
a running force protocol on the characteristic function of work.
The results are illustrated by means of the Landau-Zener(-St\"uckelberg-Majorana) model.
In the limit of continuous measurements the quantum Zeno effect
suppresses any unitary dynamics.
It is demonstrated that the characteristic function of work is
the same as for an adiabatic protocol when the continuously
measured quantity coincides with the Hamiltonian governing the
unitary dynamics of the system in the absence of measurements.
\end{abstract}


\maketitle
\section{Introduction}
The Jarzynski equality and the fluctuation theorems are general
and surprisingly robust exact relations of non-equilibrium
thermodynamics.
The validity of these relations was rigorously established
irrespective of the speed of the externally applied
forcing protocol for driven classical
\cite{Bochkov77SPJETP45,Jarzynski97PRL78,Crooks99PRE60} and
quantum systems \cite{Tasaki00arxiv,Talkner07JPA40},
staying either in complete isolation or in weak
\cite{Talkner09JSM09} or even strong contact with an environment
\cite{Jarzynski04JSM04,Campisi09PRL102}. See
also the reviews \cite{Esposito09RMP81,Campisi11RMPXX} and
references therein.
In this work we will focus on the Tasaki-Crooks work fluctuation
theorem, reading:
\begin{equation}
 \frac{p_F(w)}{p_B(-w)}=e^{\beta(w-\Delta F)}\, ,
\label{eq:TCFT}
\end{equation}
where $p_F(w)$ is the probability density function (pdf) of work
performed by a forcing protocol denoted as $\lambda_t$. This 
forcing acts on the system between the times $t=0$ and $t=\tau$. 
Accordingly, $p_B(w)$ is the pdf of 
work performed on the system when the backward (B) protocol
$\lambda_{\tau-t}$ describes the forcing of the system. 
The forward and backward processes start in the Gibbs equilibrium
states at the same inverse temperature $\beta$ and at the initial
parameter values $\lambda_0$ and $\lambda_\tau$, respectively.
The free energy difference between these two states is denoted by
$\Delta F$.

In the derivations \cite{Tasaki00arxiv,Talkner07JPA40} of the 
quantum work fluctuation theorem, Eq. (\ref{eq:TCFT}),
the energy of the system is measured at times $t=0$ and $t=\tau$,
and the work $w$ is determined by the difference of the obtained
eigenvalues.
Recently, we showed that the Tasaki-Crooks work fluctuation
theorem, Eq. (\ref{eq:TCFT}), as well as other quantum fluctuation
theorems, remain unaffected for other scenarios than this
two-measurement scheme: the ratio of forward and backward pdf's
in Eq. (\ref{eq:TCFT}) stays unaltered even if further projective
quantum measurements of any sequence of arbitrary observables are
performed while the protocol is in action \cite{Campisi10PRL105}.
In Sec. \ref{sec:CharFun} we provide an alternative proof of this
result, based on the calculation of the characteristic function
of work of a driven quantum system whose dynamics is interrupted
by projective quantum measurements.

Based on the fact that the process of measurement of a quantum
system is rather invasive due to the collapse of the wave
function,
it has been argued \cite{Campisi10PRL105,Morikuni10arxiv} that, 
although the value of the ratio of backward and forward pdf's in
Eq. (\ref{eq:TCFT}) remains unchanged, additional measurements
affect the values of the individual work pdf's.
One must expect that, in a driven-measured quantum system, the
work done is not only determined by the interaction with the
manipulated external field $\lambda$, but also by the
measurements themselves, which are physically realized by a 
measurement apparatus.
In Sec. \ref{sec:example} we calculate the statistics of work in
a prototypical model of driven quantum system, namely the
Landau-Zener(-St\"{u}ckelberg-Majorana)
\cite{Landau32PZS2,Zener32PRSA137,Stueckelberg32HPA5,
Majorana32NC9} model, and illustrate how it is influenced by
projective quantum measurements. We will also draw the attention
on interesting features related to the quantum Zeno effect
appearing when the measurements become very frequent. 
As we will see, if the observable that is measured at time $t$ is
the Hamiltonian $H(t)$, then in the Zeno limit the work
characteristic function approaches the
same expression as for an adiabatic protocol with no intermediate
measurements.

Sec. \ref{sec:discussion} closes the paper with some concluding
remarks.

\section{\label{sec:CharFun}Fluctuation Theorems for
driven-measured quantum systems}

We consider a quantum system that, in the time span $[0,\tau]$,
is thermally isolated, and interacts with the external world only
through a mechanical coupling to a time dependent external force
field and a measurement apparatus.
The information regarding this interaction is encoded in a
protocol which we denote as
\begin{equation}
\sigma=\{H(\lambda_t),(t_i,\mathcal A_i)\}.
\end{equation}
It specifies a) the system Hamiltonian $H(\lambda_t)$ at each
time $t$ in terms of the external forces $\lambda_t$, and b) the
times $t_i\in (0,\tau)$, $i=1 \dots N$,
at which measurements of the observables $\mathcal A_i$ occur.
For the sake of simplicity, in the following we will
adopt the notation $H(t)$ for $H(\lambda_t)$.
As in the two-measurement scheme
\cite{Tasaki00arxiv,Talkner07JPA40} of the Tasaki-Crooks theorem,
we assume that, besides the $N$ intermediate measurements of 
$\mathcal A_1, \dots \mathcal A_N$,
the  energies determined by $H(0)$ and $H(\tau)$, are measured at
times $t_0=0$ and $t_f=\tau$. For simplicity we then set $f=N+1$
and \begin{align}
\mathcal A_0=H(0) \qquad \mathcal A_{f}=H(\tau) \, .
\end{align}
The remaining observables, $\mathcal A_1, \dots \mathcal A_N$,
represented by hermitean operators are required to have discrete
spectra, but otherwise can be chosen arbitrarily.

For times $t\leq0$ the system is assumed to stay in the Gibbs
thermal state at inverse temperature $\beta$:
\begin{equation}
\rho_0^\beta=e^{-\beta H(0)}/Z_0 \, ,
\end{equation}
where
\begin{equation}
Z_0=\Tr e^{-\beta H(0)}
\end{equation}
is the canonical partition function. 

We denote the orthogonal eigenprojectors of the observable
$\mathcal A_i$ by $\Pi_k^i$. These satisfy the eigenvalue
equations:
\begin{equation}
\mathcal A_i \Pi_k^i= \alpha_k^i \Pi_k^i\, ,
\label{eq:Ai}
\end{equation}
where $\alpha_k^i$ are the eigenvalues of $\mathcal A_i$. Hence,
with the above choice of the first and last measured observables,
 $\alpha_k^0$ and  $\alpha_k^f$ represent the instantaneous
eigen-energies of the system at $t=0$ and $t=\tau$, respectively.

In the following we compute the conditional probability
$p_\sigma(m,\tau|n,0)$ to find the eigenvalue $\alpha_m^f$ 
obtained 
in the last measurement at time $\tau$, provided that the
eigenvalue $\alpha_n^0$ was the result of the first measurement,
under those conditions that are specified by the protocol
$\sigma$. 
We begin our discussion considering only one intermediate
measurement ($N=1$) of the observable $\mathcal A_1$.

According to the von Neumann postulates, immediately after the
eigenvalue $\alpha_n^0$ is measured at $t=0$, the system density
matrix becomes:
\begin{equation}
\rho_n(0^+)=\Pi_n^0\rho_0^\beta \Pi_n^0/p_n\, ,
\end{equation}
where
\begin{equation}
p_n= \Tr\, \Pi_n^0\rho_0^\beta /Z_0= e^{-\beta \alpha_n^0}/Z_0
\end{equation}
is the probability to find the system initially in the state with
energy $\alpha_n^0$.
Since we assumed that the system is thermally isolated, it
subsequently evolves until time $t_1$ according to the unitary
time evolution $U_{t_1,0}$ that is governed by the
Schr{\"o}dinger equation:
\begin{equation}
i\hbar \partial_t U_{t,0} = H(t) U_{t,0}, \quad
U_{0,0}=\mathbb{1} \, .
\end{equation}
Thus, immediately before the measurement of $A_1$, occurring at
$t_1$, the density matrix is:
\begin{equation}
\rho_n(t_1^-)= U_{t_1,0} \rho_n(0^+) U_{t_1,0}^\dagger\, ,
\end{equation}
and the subsequent measurement projects it into
\begin{equation}
\rho_n(t_1^+)=\sum_r \Pi_r^1 U_{t_1,0} \rho_n(0^+)
U_{t_1,0}^\dagger \Pi_r^1 \, .
\end{equation}
Likewise, just before the measurement of $\mathcal A_f=H(\tau)$
at time $\tau$, the density matrix becomes
\begin{equation}
\rho_n(\tau^-)= U_{\tau,t_1} \rho_n(t_1^+) U_{\tau,t_1}^\dagger
\, ,
\end{equation}
and the probability that the outcome of the measurement of
$\mathcal A_f=H(\tau)$ at time $\tau$ is $\alpha_m^f$ is:
\begin{equation}
p_\sigma(m,\tau|n,0)=\Tr\, \Pi_m^f U_{\tau,t_1} \rho_n(t_1^+)
U_{\tau,t_1}^\dagger \, .
\end{equation}

Finally, the pdf of work $w$, $P_\sigma(w)$, is obtained as the
sum of the joint probability $p_\sigma(m,\tau|n,0) p_n$
restricted to 
\begin{equation}
w= \alpha_n^f-\alpha_n^0 \, ,
\end{equation}
and hence becomes
\begin{equation}
P_\sigma(w)=\sum_{n,m} \delta[w-\alpha_n^f-\alpha_n^0]
p_\sigma(m,\tau|n,0)p_n \, .
\label{eq:P(w)}
\end{equation}

\subsection{The characteristic function of work}
Next we focus on the characteristic function of work
$G_\sigma(u)$, given by the Fourier transform of the work pdf
$P_\sigma(w)$:
\begin{equation}
G_\sigma(u) = \int dw P_\sigma(w)e^{iuw} \, .
\label{eq:CF}
\end{equation}
Substituting (\ref{eq:P(w)}) into (\ref{eq:CF}) we obtain for the
characteristic function:
\begin{align}
G_\sigma(u) &= \sum_{m,n} e^{iu[\alpha_m^{f} -\alpha_n^{0}]}
p_\sigma(m,\tau|n,0) e^{-\beta \alpha_n^{0}}/Z_0\nonumber \\
&= \sum_{m,n,r} e^{iu[\alpha_m^f -\alpha_n^{0}]}\Tr\, \Pi_m^f
U_{\tau,t_1}\Pi_r^1 U_{t_1,0}\nonumber \\
&\quad \times  \Pi_n^0 \rho_0^\beta
\Pi_n^0 U^\dagger_{t_1,0} \Pi_r^1
U^\dagger_{\tau,t_1}\nonumber \\
&=\sum_r \Tr \, e^{iu H(\tau)} U_{\tau,t_1} \Pi_r^1 U_{t_1,0}
e^{-iu H(0)} \rho_0^\beta \nonumber \\
&\quad \times U^\dagger_{t_1,0}\Pi_r^1 U^\dagger_{\tau,t_1}
\nonumber\\
&=\Tr \left [e^{iu H(\tau)} \right ]_\sigma e^{-iu
H(0)}\rho_0^\beta \, ,
\label{G}
\end{align}
where $[X]_\sigma$ denotes the time evolution of an operator $X$
from $t=0$ to $t=\tau$  in presence of the protocol $\sigma$,
that implies a unitary evolution governed by $H(t)$ interrupted
by a measurement of an observable  $\mathcal A_1$ at time
$t=t_1$. It takes the form
\begin{equation}
\left [ X \right ]_\sigma = \sum_r U^\dagger_{t_1,0}
\Pi_r^1 U^\dagger_{\tau,t_1} X U_{\tau,t_1} \Pi_r^1
U_{t_1,0}, \quad N=1 \, .
\label{Ht}
\end{equation}
In the case of a protocol $\sigma$ containing $N$ interrupting
measurements 
the formal expression of the characteristic function is the same
as for one interrupting measurement, Eq. (\ref{G}) with the time
evolution $[X]_\sigma$ given by
\begin{align}
[ X ]_\sigma=& \sum_{r_1 \dots r_N}
U^\dagger_{t_1,0} \Pi_{r_1}^1  U^\dagger_{t_2,t_1}  
\Pi_{ r_2}^2 \dots \nonumber\\
  &\dots  U^\dagger_{t_N,t_{N-1}}  \Pi_{ r_N}^N 
U^\dagger_{\tau,t_N}X U_{\tau,t_N} \Pi_{ r_N}^N  U_{t_N,t_{N-1}}
\dots \nonumber \\
&  \ldots \Pi_{ r_2}^2 U_{t_2,t_1} \Pi_{r_1}^1 U_{t_1,0} \, .
\label{Htt}
\end{align}

\subsection{The Jarzynski equality}
Putting $u=i \beta$,
one recovers the Jarzynski equality:
\begin{align}
G_\sigma(i \beta)&=\langle e^{-\beta w}\rangle_\sigma= \Tr
\left [ e^{-\beta H(\tau)} \right ]_\sigma e^{\beta H(0)}
\rho_0^\beta\nonumber \\
&= Z_0^{-1}\, \Tr\, \left [ e^{-\beta H(\tau)} \right
]_\sigma \nonumber \\
&= Z_0^{-1}\, \Tr\,  e^{-\beta H(\tau)}=Z_f/Z_0 =e^{-\beta \Delta
F}\, ,
\label{eq:JEfromG}
\end{align}
because $ \Tr [X]_\sigma=
\Tr X $ for any trace class operator $X$ \cite{Holevo01Book}.
This follows from
the cyclic invariance of the trace, the unitarity relation
$U_{t,s}^\dagger U_{t,s}=\mathbb{1}$, and the completeness of the
projection operators $\sum_r \Pi_r^i=\mathbb{1}$. Here $Z_f=
\Tr\, e^{-\beta H(\tau)}$ denotes the partition function of the
Gibbs state at the initial temperature and final parameter
values:
\begin{equation}
\rho_f^\beta=e^{-\beta H(\tau)}/Z_f \, ,
\label{eq:final-rho}
\end{equation}
and $\Delta F=-\beta^{-1}\ln Z_0/Z_f$ is the difference of the free
energies of the thermal equilibrium states $\rho_f^\beta$ and
$\rho_0^\beta$.

The symbol $\langle\cdot\rangle_\sigma$ denotes an average with
respect to the work pdf $P_\sigma(w)$. Eq. (\ref{eq:JEfromG})
says that the Jarzynski equality holds irrespective of the
details of the interaction protocol $\sigma$. 
Independent of number and nature of the measured observables as
well as strength, speed, and functional form the driving force 
the average exponentiated work is solely determined by  the free
energy difference \cite{Campisi09PRL102}.

\subsection{The work fluctuation theorem}
Besides the Jarzynski equality also 
the Tasaki-Crooks theorem is robust under repeated quantum
measurements.
In the
presence of many intermediate measurements it reads:
\begin{equation}
P_{\sigma}(w)={P_{\tilde{\sigma}}(-w)}e^{\beta (w-\Delta F)} \, ,
\label{eq:TC}
\end{equation}
where the tilde ($\tilde{\sigma}$) indicates the temporal
inversion of the protocol $\sigma$, that is:
\begin{equation}
\tilde{\sigma}= \{H(\tau-t),(\tau-t_i,\mathcal A_i)\}\, .
\end{equation}
Hence, $\tilde{\sigma}$ specifies the succession of force values
and measurements in the reversed order, specifically with the
measurement of the observables $\mathcal A_i$ at times
$\tau-t_i$. In particular it implies that at time $0$, the
observable $\mathcal A_f=H(\tau)$ and at time $\tau$, the
observable $\mathcal A_0=H(0)$ are measured.
Accordingly, the initial state of the backward process is given
by 
the Gibbs state $\rho_f^\beta$, Eq. (\ref{eq:final-rho}), i.e.
the system is at equilibrium with inverse temperature $\beta$ and
 force value $\lambda_\tau$.

Eq. (\ref{eq:TC}) holds under the assumptions that both $H(t)$
and $\mathcal A_i$ commute with the quantum mechanical
anti-unitary time reversal operator $\Theta$
\cite{Messiah62Book}. That is, for all $t\in[0,\tau]$, and $i=0
\dots f$, we assume
\begin{align}
H(t)\Theta&=\Theta H(t) \label{HTTH}\\
\mathcal A_i \Theta&=\Theta \mathcal A_i  \, .
\label{eq:ThetaA=Atheta}
\end{align}

We prove Eq. (\ref{eq:TC}) for the simplest case of a single
intermediate measurement, $N=1$.
The generalization to many measurements is straightforward. From
Eq. (\ref{G}) we have
\begin{align}
G_\sigma(u)
=&\Tr \sum_r U^\dagger_{t_1,0}\Pi_r^1
U^\dagger_{\tau,t_1}  e^{iu H(\tau)}  \nonumber \\
& \times U_{\tau,t_1} \Pi_r^1
U_{t_1,0} e^{i(i\beta-u) H(0)} /Z_0 \, .
\label{eq:G2}
\end{align}
Introducing the
notation $\widetilde U_{t',t}$ for the time evolution governed
by $\widetilde H(t)\equiv H(\tau-t)$, the
backward characteristic function of work can be written as:
\begin{align}
G&_{\tilde{\sigma}}(u) = \Tr \sum_r
\widetilde U^\dagger_{\tau-t_1,0}  \Pi_r^1 \widetilde
U^\dagger_{\tau,\tau-t_1}
 e^{iu \widetilde H(\tau)} \nonumber \\
& \qquad \times \widetilde U_{\tau,\tau-t_1}\Pi_r^1\widetilde 
U_{\tau-t_1,0}  e^{-iu \widetilde H(0)} e^{-\beta
H(\tau)}/Z_f \nonumber \\
=& \Tr \sum_r \Theta \left[  \Theta^\dagger \widetilde
U^\dagger_{\tau-t_1,0} \Theta \Theta^\dagger \Pi_r^1\Theta
\Theta^\dagger
\widetilde U^\dagger_{\tau,\tau-t_1}
\Theta \right.\nonumber \\
&\times \Theta^\dagger e^{iu \widetilde H(\tau)}
\Theta \Theta^\dagger \widetilde U_{\tau,\tau-t_1} 
\Theta \Theta^\dagger\Pi_r^1\Theta\nonumber \\
&\left.
\times  \Theta^\dagger\widetilde  U_{\tau-t_1,0}
\Theta \Theta^\dagger e^{-i(u+i\beta) \widetilde H(0)}
\Theta\right]\Theta^\dagger/Z_f \, ,
\label{Gtilde}
\end{align}
where we used the antiunitarity
$\Theta\Theta^\dagger=\mathbb{1}$.
From Eq. (\ref{eq:ThetaA=Atheta})  it follows that all
eigenprojection operators commute with the time-reversal
operator, i.e. 
$\Pi_r^1\Theta=
\Theta \Pi_r^1$. The time reversal invariance, expressed by Eq.
(\ref{HTTH}) 
implies $ \Theta^\dagger  e^{s H(t)} \Theta= e^{s^*
H(t)}$ for any $C$-number $s$. Further, microreversibility of
driven systems
\cite{Andrieux08PRL100,Campisi11RMPXX} implies
\begin{align}
\Theta^\dagger \widetilde U_{\tau-t,0} \Theta=U_{t,\tau}
\nonumber \\
\Theta^\dagger \widetilde U_{\tau,\tau-
t}\Theta= U_{0,t}\, .
\label{eq:q-microrev}
\end{align}
Using these relations and recalling that, for any trace class
operator $X$,  $\Tr\, \Theta X  \Theta^\dagger=\Tr X^\dagger$, 
one ends up with:
\begin{align}
G_{\tilde{\sigma}}(u)
=&\Tr \sum_r e^{i(i\beta-u) H(\tau)} U_{\tau,t_1} \Pi_r^1
U_{t_1,0} \nonumber \\
& \times e^{iu H(0)} U^\dagger_{t_1,0}\Pi_r^1
U^\dagger_{\tau,t_1}/Z_f \, .
\end{align}
By comparison with Eq. (\ref{eq:G2}), we finally find
\begin{equation}
Z_fG_{\tilde{\sigma}}(i\beta -u)=Z_0 G_\sigma(u) \, ,
\end{equation}
hence, by means of an inverse Fourier transform the searched
fluctuation theorem (\ref{eq:TC}).

We thus have proved that the fluctuation theorem of Tasaki-Crooks
remains unchanged if additionally to the measurements of energy
at time $t=0$ and $t=\tau$, intermediate measurements of time
reversal invariant observables $\mathcal A_i$ are performed at
times $t_i$, provided 
the order of measurements in the backward protocol is properly
changed in accordance with 
the corresponding 
times $\tau-t_i$.

\section{\label{sec:example}Example}
As mentioned in the introduction, one expects measurements to
strongly influence the pdf of work although the Jarzynski
equality and the Tasaki-Crooks relation are insensitive
to intermediate measurements.
To illustrate this point in more detail, we consider the example
of the
Landau-Zener(-St\"{u}ckelberg-Majorana)
\cite{Landau32PZS2,Zener32PRSA137,Stueckelberg32HPA5,
Majorana32NC9} model described by the Hamiltonian:
\begin{equation}
 H(t)= \frac{vt}{2}\sigma_z +\Delta \sigma_x\, .
\label{eq:H-LZ}
\end{equation}
It governs the dynamics of a two-level quantum system whose 
energy separation, $vt$, varies linearly in time, and whose states 
are coupled via the interaction energy $\Delta$.
Here, $\sigma_x$ and $\sigma_z$ denote Pauli matrices.

The Landau-Zener model is one of
the few time-dependent quantum mechanical problems  that have an
analytic solution. The elements of the $2\times 2$ unitary time
evolution matrix $U_{t,s}$ can be expressed in terms of parabolic
cylinder functions \cite{Vitanov99PRA59}. The instantaneous
eigenvalues of $H(t)$ are:
\begin{equation}
 E_n^t= (n-1/2)\sqrt{v^2t^2+4\Delta^2}\, , \quad n=0,1\, .
\end{equation}
Since these energies are symmetric with respect to an inversion
about $t=0$
we choose the initial and final times $t_0$ and $t_f$, as
$-\tau/2$ and $\tau/2$, respectively, instead of $0$ and $\tau$,
as in the previous discussion.
\begin{figure}[b]
\includegraphics[width=7.5cm]{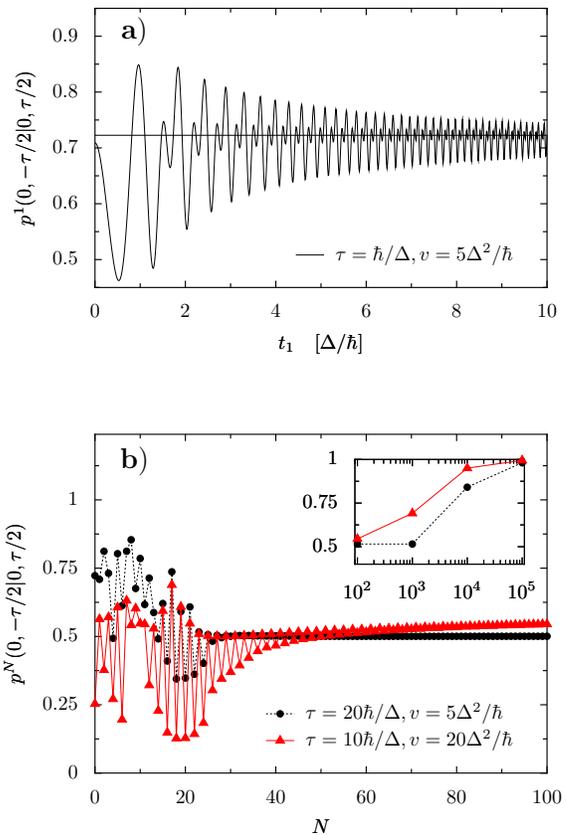}
\caption{(Color online) Survival probability of the Landau-Zener system, Eq.
(\ref{eq:H-LZ}), interrupted by measurements.
a) Survival probability
as a function of the time $t_1$ of a single energy measurement of
$H(t_1)$.
The parameter values are $\tau=20 \hbar /\Delta$,
 $v=5 \Delta^2/\hbar$. The horizontal line is
the survival probability
$p(0,-\tau/2|0,\tau/2)$ with no intermediate measurements.
b) Survival probability as a function of the number
$N$
of equally spaced intermediate measurements of $H(t)$. The
parameter values are $\tau=20 \hbar /\Delta$, 
$v=5 \Delta^2/\hbar$ (black circles), and 
$\tau=10 \hbar /\Delta$, $v=20 \Delta^2/\hbar$ (red triangles).
Inset: the survival probabilities approach the value $1$ in the
Zeno-limit.
}
\label{fig:Fig}
\end{figure}

Figure \ref{fig:Fig}a) depicts the survival
probability $p^1(0,\tau/2|0,-\tau/2)$ as a function of the
instant $t_1$ of a single intermediate measurement of $H(t_1)$
for positive times $t_1$ and fixed length $\tau$ of the protocol.
As an even function of $t_1$ this also specifies
$p^1(0,\tau/2|0,-\tau/2)$ for negative $t_1$.
The straight horizontal line shows the value of
the survival probability $p(0,\tau/2|0,-\tau/2)$ 
without intermediate measurement. It is obvious that the
intermediate measurement in general alters the survival
probability thus affecting the work pdf. For example, the average
work,
\begin{equation}
\langle w
\rangle= 2 E_1^{\tau/2}\tanh[\beta
E_1^{\tau/2}][1-p(0,\tau/2|0,-\tau/2)]\, ,
\end{equation}
evidently changes when $p(0,\tau/2|0,-\tau/2)$ is replaced by
$p^{1}(0,\tau/2|0,-\tau/2)$. The same can be said for the
standard
deviation of work that reads 
\begin{equation}
\langle \Delta w^2\rangle =
(2E_1^{\tau/2})^2[1-p(0,\tau/2|0,-\tau/2)]-\langle w \rangle^2 \,
.
\end{equation}
Notably, the introduction of an intermediate measurement may 
lower the average work.

Figure \ref{fig:Fig}b) shows the $N$ dependence
of the survival probability
$p^N(0,\tau/2|0,-\tau/2)$ for $N$ equally spaced
intermediate measurements of $H(t)$, and two sets of model
parameters. Oscillatory behavior is observed for
small values of $N$, while as $N$ increases the asymptotic value
$1$ is approached, see the inset. 
When the measurement frequency is high enough the
unitary dynamics between subsequent measurements becomes
increasingly suppressed until the dynamics is completely frozen
and consequently the 
survival probability reaches the asymptotic value 1 for $N \to
\infty$.
This phenomenon is known as the quantum Zeno effect
\cite{Misra77JMP18,Peres80AJP48,Koshino05PREP412,Facchi08JPA41}.
We investigate it further in the following section.

\subsection{Quantum Zeno Effect}
To formally elucidate the quantum Zeno effect observed in
this particular example, and under more general
conditions as well, we analyze the form of the characteristic function
given by Eq. (\ref{G}) in the limit of infinitely many
measurements of energy.
We approach this limit by considering a
finite number of $N$ intermediate measurements of energy that 
take place at equally spaced instants 
\begin{equation}
 t_k = t_0 + k \varepsilon \, ,\qquad k=1\dots N\, ,
\end{equation}
 where
$\varepsilon=(t_f-t_0)/(N+1)$ denotes the time elapsing between two
subsequent measurements. We denote the corresponding protocol
with symbol $\nu$, that is:
\begin{equation}
\nu=\{H(t),(t_k,H(t_k))\}\, .
\end{equation}
Since we are interested in the limiting case of infinitely many
measurements, we may choose $N$ sufficiently large such that the
Hamiltonian between two subsequent measurements can safely be
approximated by  its value at the later measurement, i.e. $H(t)
\approx H(t_k)$ for $t \in (t_{k-1},t_k)$ with $t_k= t_0 + k \varepsilon$
being the instant of the $k$th measurement.
The time evolution within such a short period then becomes
\begin{equation}
U_{t_k,t_{k-1}}\simeq e^{-iH(t_k)\varepsilon/\hbar} \:.
\end{equation}  
According to Eq. (\ref{Htt}) the time evolution
$U^\dagger_{t_k,t_{k-1}}$ acts on the projection operator
$\Pi_{r_k}(t_k)$ resulting in the phase-factor $e^{i
\alpha_{r_k}(t_k) \varepsilon/\hbar}$ while the complex conjugate phase
factor is obtained from the product $\Pi_{r_k}(t_k)
U_{t_k,t_{k-1}}$ which appears right of the operator $X$
\footnote{In  slight deviation from our previous notation, Eq.
(\ref{eq:Ai}), we denote the eigenvalues of $\mathcal
A_k=H(t_k)$, by $\alpha_r(t_k)$, and the respective projection
operators by $\Pi_r(t_k)$.}.
 Hence, these factors cancel each other and one obtains for the
$\nu$-propagated exponential operator $X= e^{iuH(\tau)}$ the
expression
\begin{equation}
\begin{split}
\left [ e^{iu H(\tau)} \right ]_{\nu} &= \sum_{r_1,r_2,\dots
r_N} \Pi_{r_1}(t_1) \Pi_{r_2}(t_2) \ldots \\
& \quad \times  \Pi_{r_N}(t_N) e^{i u H(\tau)} \Pi_{r_N}(t_N)  
\ldots \\
& \quad \times  \Pi_{r_2}(t_2)\Pi_{r_1}(t_1)\, .
\end{split}
\label{XZ}
\end{equation}
Assuming first that $H(t)$ is non-degenerate at any time $t \in
(t_0,t_f)$ all projection operators are of the one-dimensional
form $\Pi_{r_k}(t_k) = |\psi_{r_k}(t_k) \rangle\langle
\psi_{r_k}(t_k)| $ with instantaneous eigenfunctions
$|\psi_{r_k}(t_k) \rangle$ of the Hamiltonian $H(t_k)$ .  The
labeling of the instantaneous eigenstates can be arranged in such
a way that the eigenvalues $\alpha_k(t)$ are continuous functions
of time. In other words, each adiabatic energy branch is labeled
by an index $r$. The scalar products of eigenfunctions at
neighboring times then deviate  from a Kronecker delta by terms
of the order $\varepsilon^2$ which consequently can be neglected, i.e.,
states on different adiabatic energy branches at neighboring
times are almost orthogonal
\begin{equation}
\langle \psi_{r_k}(t_k)|\psi_{r_{k+1}}(t_{k+1}) \rangle =
\delta_{r_k,r_{k+1}} + \mathcal{O}(\varepsilon^2)\, .
\label{sp}
\end{equation}
Herewith the left hand side of Eq. (\ref{XZ}) simplifies to read
in the limit of infinitely many dense measurements 
\begin{equation}
\left [ e^{iu H(\tau)} \right ]_{\nu} = \sum_r |\psi_r(t_0)
\rangle \langle \psi_r(t_f)|e^{iu H(t_f)} |\psi_r(t_f)
\rangle\langle\psi_r(t_0)|\, .
\label{XP}
\end{equation}
For the generating function this yields 
\begin{equation}
G_{\nu}^{\infty}(u) = \sum_r e^{iu [\alpha_r(t_f) - \alpha_r(t_0)]} e^{-\beta
\alpha_r(t_0)}/Z_0\, .
\label{GZ}
\end{equation}
We therefore find that in the Zeno-limit the characteristic
function of work and accordingly the pdf of work coincide with
the respective expressions obtained for an adiabatic protocol in
the absence of any intermediate measurement. 

The same line of reasoning also applies for the case that an
observable that does not change in time is repeatedly measured,
i.e., for a protocol
\begin{equation}
 \mu=\{H(t),(t_k,\mathcal A)\} \, .
\end{equation}
For such an
observable $\mathcal{A}$
with eigenprojection operators $\Pi_r$ one then obtains in the
Zeno limit
for the characteristic function
\begin{equation}
G^{\infty}_{\mu}(u) = \Tr \sum_r \left (\Pi_r e^{iu
H(t_f)} \Pi_r\right ) e^{-iuH(t_0)} e^{-\beta H(t_0)}\, .
\end{equation}

\subsubsection{Landau-Zener}
In the case of the Landau-Zener problem studied in the previous 
section, Eq. (\ref{eq:H-LZ}), the eigenvalues on each adiabatic 
branch coincide at the beginning and the end of the protocol and
therefore the coefficients of $u$ in the exponential terms on the
right hand side of Eq. (\ref{GZ}) vanish leading to
\begin{equation}
G^{\infty}_{\nu} (u)=1\, .
\label{GLZZ}
\end{equation}
This leads to the expected result that under permanent 
observation of the Hamiltonian no work is done, i.e.,
\begin{equation}
P^{\infty}_{\nu}(w) = \delta(w)\, .
\end{equation}
The Zeno effect sets in when the measurement frequency is so
large that any unitary evolution within the time leaps between
two subsequent measurements can be neglected, hence for
$\varepsilon \sqrt{v^2 \tau^2/4 + 4 \Delta}/\hbar \ll 1$, or 
equivalently if $N\gg \tau \sqrt{v^2 \tau^2/4 + 4 \Delta}/\hbar$.
With the parameters used in Fig. \ref{fig:Fig}b), the Zeno effect 
sets in for $N \gg 10^3$, as shown in the inset of Fig. \ref{fig:Fig}b).

In the case of degeneracy, Eq. (\ref{GZ}) continues to hold if
the instantaneous energy branches are labeled such that the
corresponding instantaneous eigenvalues smoothly vary along each
branch. For the  Landau-Zener problem a degeneracy happens at
$t=0$ if the coupling strength $\Delta$ vanishes. One then finds
the characteristic function
\begin{equation}
G^{\infty}_{\nu} (u) = \frac{\cosh [(\beta + 2 i
u)v\tau/2]}{\cosh (\beta v\tau/2)}\qquad [\Delta=0]\, ,
\label{GLZZ0}
\end{equation}  
yielding for the pdf of work
\begin{equation}
P_{\nu}^{\infty}(w) = p_- \delta(w-v\tau) + p_+
\delta(w+v\tau)\qquad [\Delta=0]\, ,
\end{equation}
where $p_{\pm} = e^{\mp \beta v \tau/2}/[2 \cosh (\beta \tau/2)]$
denote the thermal populations of the ground- and the excited
states at the initial time $t=-\tau/2$, respectively.

In the case of countinuous measurement of the Pauli operators
$\sigma_i$, $i=x,y,z$, the driving-measuremet protocols are
\begin{equation}
 \mu_i = \{H(t),(t_k,\sigma_i)\},\quad i=x,y,z \, .
\end{equation}
The characteristic functions of work then become
\begin{equation}
\begin{split}
G^{\infty}_{\mu_x}(u) &=\frac{1}{2} \left (
1-\frac{\Delta^2}{q^2} \right ) \left (p_- e^{2i q u} + p_+
e^{-2i q u} \right )\\
&\quad +\frac{1}{2} \left ( 1+\frac{\Delta^2}{q^2} \right )     
\\
G^{\infty}_{\mu_y}(u) &= \frac{1}{2} \left ( p_- e^{2iq u} +
p_+ e^{-2i q u}  +1 \right )  \\
G^{\infty}_{\mu_z}(u) &=\frac{1}{2} \left (
1+\frac{\Delta^2}{q^2} \right ) \left (p_- e^{2i q u} + p_+
e^{-2i q u} \right )\\
&\quad +\frac{1}{2} \left ( 1-\frac{\Delta^2}{q^2} \right ) \, , 
   \\ 
\end{split}
\end{equation}
where $p_\pm$ denotes the populations of the exited and the
ground states of the initial Hamiltonian, respectively, given by
\begin{equation}
p_\pm =e^{\mp \beta q}/\left ( e^{\beta q} + e^{-\beta q} \right
)\, ,
\end{equation}
with 
\begin{equation}
q = \sqrt{(v \tau/2)^2+\Delta^2}\:.
\end{equation}
The work pdf follows then by means of an inverse Fourier
transform to read
\begin{equation}
\begin{split}
p^\infty_{\mu_x} &= \frac{1}{2} \left (1-\frac{\Delta^2}{q^2}
\right )  ( p_- \delta(w-2 q) + p_+ \delta(w+2 q) )\\
&\quad  + \frac{1}{2} \left (1+\frac{\Delta^2}{q^2} \right )
\delta(w)\\
p^\infty_{\mu_y} &= \frac{1}{2}  ( p_- \delta(w-2 q) + p_+
\delta(w+2 q) +\delta(w))\\
p^\infty_{\mu_z} &= \frac{1}{2} \left (1+\frac{\Delta^2}{q^2}
\right )  ( p_- \delta(w-2 q) + p_+ \delta(w+2 q) )\\
&\quad  + \frac{1}{2} \left (1-\frac{\Delta^2}{q^2} \right )
\delta(w)\, .\\ 
\end{split}
\end{equation}

\section{\label{sec:discussion}Discussion}
We provided an alternative proof of the
robustness of the Tasaki-Crooks fluctuation theorem 
against repeated measurements \cite{Campisi10PRL105}.
The present proof however is more general in regard to the fact
that it allows for possibly degenerate eigenvalues of the
measured observables. Further it was found that the
characteristic function of work assumes the form of a quantum
two-time correlation function of the exponentiated initial
Hamiltonian, and the exponentiated final Hamiltonian in the
Heisenberg picture generated by the unitary evolution interrupted
by quantum collapses.

Our proof  keeps holding also if
quantum measurements are done with respect to 
positive operator valued measures (POVM) \cite{Peres93Book}
which are less invasive than those described by  von Neumann
projection valued measures. For the case
of measurements with respect to a POVM, 
\begin{equation}
 \sum_r M_r
M_r^\dagger=\mathbb{1}
\label{M1}
\end{equation}
the interrupted evolution
$[X]_\sigma$ is obtained from Eq. (\ref{Htt}) with
the projection operators $\Pi_{r}$ on the left hand side of $X$
being replaced by the weak measurement operators $M_r$, and
those right of $X$ by the adjoint operators $M_r^\dagger$. Due to
Eq. (\ref{M1}), $\Tr [X]_\sigma = \Tr X $ continues to hold
implying the validity of the Tasaki-Crooks relation, Eq.
(\ref{eq:TC}), for weak measurement of time reversal invariant
observables with $M_k\Theta=\Theta M_k$.

In proving Eq. (\ref{eq:TC}) we assumed that the Hamiltonian and
the measured observables commute with the time reversal operator.
This assumption may be relaxed and the Tasaki-Crooks fluctuation
theorem continues to hold 
if the backward protocol is defined as:
\begin{equation}
\tilde{\sigma}= \{\Theta H(\tau-t)
\Theta^\dagger,(\tau-t_i,\Theta \mathcal A_i \Theta^\dagger)\} \,
.
\end{equation}

Notwithstanding the robustness of the Tasaki-Crooks theorem, we 
found,
in accordance with the intuitive expectation, that
the interruption of the dynamics of a driven quantum system by
means of projective measurements alters the statistics of work
performed on the system. We illustrated the influence of
measurements on the work distribution by the example of the
Landau-Zener(-St\"{u}ckelberg-Majorana) model. We noticed that
depending on the positions in time or  frequency of one or more
interrupting measurements, the average work $\langle w \rangle$
may be lowered, Fig.~\ref{fig:Fig}.

\citeauthor{Schmiedl09JSM09} \cite{Schmiedl09JSM09} studied the problem of designing 
optimal protocols that minimize the average work spent during
the forcing, without considering intermediate measurements.
In order to further minimize the average 
work, one might expand the optimization parameter space and 
include the possibility of performing intermediate measurements.
This opens for the possibility of a much greater control over the 
energy flow into the system.

In the limit of very frequent measurements the unitary dynamics
becomes completely suppressed due to the quantum Zeno effect. If
at each measurement time the instantaneous Hamiltonian is
measured, then in the Zeno limit the work characteristic function
approaches the same form that it would assume if the protocol was
adiabatic.

The result of an intermediate measurement may be used to alter
the 
subsequent force protocol. In this way a feedback control can be
implemented
for classical \cite{Sagawa10PRL104} as well as for quantum
systems  \cite{Morikuni10arxiv}. In both cases the Jarzynski
equality only holds in a modified form.

Landau-Zener dynamics and frequent quantum measurements of a
two-level
system coupled to a thermal bath were recently shown to provide
efficient means for quantum-state preparation \cite{ZuecoNJP08}
and purification \cite{Alvarez10PRL105}, respectively. These are
crucial prerequisite
for the implementation of working quantum computers. The
combined 
study of dissipative Landau-Zener dynamics with frequent 
observations, could unveil yet new practical methods for quantum
state control and manipulation.

\acknowledgements
This work was supported by the cluster of excellence
Nanosystems Initiative Munich (NIM) and the Volkswagen Foundation
(project I/80424).


\begin{thebibliography}{29}
\expandafter\ifx\csname natexlab\endcsname\relax\def\natexlab#1{#1}\fi
\expandafter\ifx\csname bibnamefont\endcsname\relax
  \def\bibnamefont#1{#1}\fi
\expandafter\ifx\csname bibfnamefont\endcsname\relax
  \def\bibfnamefont#1{#1}\fi
\expandafter\ifx\csname citenamefont\endcsname\relax
  \def\citenamefont#1{#1}\fi
\expandafter\ifx\csname url\endcsname\relax
  \def\url#1{\texttt{#1}}\fi
\expandafter\ifx\csname urlprefix\endcsname\relax\def\urlprefix{URL }\fi
\providecommand{\bibinfo}[2]{#2}
\providecommand{\eprint}[2][]{\url{#2}}

\bibitem[{\citenamefont{{Bochkov} and {Kuzovlev}}(1977)}]{Bochkov77SPJETP45}
\bibinfo{author}{\bibfnamefont{G.~N.} \bibnamefont{{Bochkov}}}
  \bibnamefont{and} \bibinfo{author}{\bibfnamefont{Y.~E.}
  \bibnamefont{{Kuzovlev}}}, \bibinfo{journal}{Zh. Eksp. Teor. Fiz.}
  \textbf{\bibinfo{volume}{72}}, \bibinfo{pages}{238} (\bibinfo{year}{1977}),
  \bibinfo{note}{[Sov. Phys. JETP \textbf{45}, 125 (1977)]}.

\bibitem[{\citenamefont{Jarzynski}(1997)}]{Jarzynski97PRL78}
\bibinfo{author}{\bibfnamefont{C.}~\bibnamefont{Jarzynski}},
  \bibinfo{journal}{Phys. Rev. Lett.} \textbf{\bibinfo{volume}{78}},
  \bibinfo{pages}{2690} (\bibinfo{year}{1997}).

\bibitem[{\citenamefont{Crooks}(1999)}]{Crooks99PRE60}
\bibinfo{author}{\bibfnamefont{G.~E.} \bibnamefont{Crooks}},
  \bibinfo{journal}{Phys. Rev. E} \textbf{\bibinfo{volume}{60}},
  \bibinfo{pages}{2721} (\bibinfo{year}{1999}).

\bibitem[{\citenamefont{Tasaki}(2000)}]{Tasaki00arxiv}
\bibinfo{author}{\bibfnamefont{H.}~\bibnamefont{Tasaki}}
  (\bibinfo{year}{2000}), \eprint{arXiv:cond-mat/0009244}.

\bibitem[{\citenamefont{Talkner and H{\"a}nggi}(2007)}]{Talkner07JPA40}
\bibinfo{author}{\bibfnamefont{P.}~\bibnamefont{Talkner}} \bibnamefont{and}
  \bibinfo{author}{\bibfnamefont{P.}~\bibnamefont{H{\"a}nggi}},
  \bibinfo{journal}{J. Phys. A} \textbf{\bibinfo{volume}{40}},
  \bibinfo{pages}{F569} (\bibinfo{year}{2007}).

\bibitem[{\citenamefont{Talkner et~al.}(2009)\citenamefont{Talkner, Campisi,
  and H{\"a}nggi}}]{Talkner09JSM09}
\bibinfo{author}{\bibfnamefont{P.}~\bibnamefont{Talkner}},
  \bibinfo{author}{\bibfnamefont{M.}~\bibnamefont{Campisi}}, \bibnamefont{and}
  \bibinfo{author}{\bibfnamefont{P.}~\bibnamefont{H{\"a}nggi}},
  \bibinfo{journal}{J. Stat. Mech.: Theory Exp.}, \bibinfo{pages}{P02025}
  (\bibinfo{year}{2009}).

\bibitem[{\citenamefont{Jarzynski}(2004)}]{Jarzynski04JSM04}
\bibinfo{author}{\bibfnamefont{C.}~\bibnamefont{Jarzynski}},
  \bibinfo{journal}{J. Stat. Mech.: Theory Exp.}, \bibinfo{pages}{P09005}
  (\bibinfo{year}{2004}).

\bibitem[{\citenamefont{Campisi et~al.}(2009)\citenamefont{Campisi, Talkner,
  and H{\"a}nggi}}]{Campisi09PRL102}
\bibinfo{author}{\bibfnamefont{M.}~\bibnamefont{Campisi}},
  \bibinfo{author}{\bibfnamefont{P.}~\bibnamefont{Talkner}}, \bibnamefont{and}
  \bibinfo{author}{\bibfnamefont{P.}~\bibnamefont{H{\"a}nggi}},
  \bibinfo{journal}{Phys. Rev. Lett.} \textbf{\bibinfo{volume}{102}},
  \bibinfo{pages}{210401} (\bibinfo{year}{2009}).

\bibitem[{\citenamefont{Esposito et~al.}(2009)\citenamefont{Esposito, Harbola,
  and Mukamel}}]{Esposito09RMP81}
\bibinfo{author}{\bibfnamefont{M.}~\bibnamefont{Esposito}},
  \bibinfo{author}{\bibfnamefont{U.}~\bibnamefont{Harbola}}, \bibnamefont{and}
  \bibinfo{author}{\bibfnamefont{S.}~\bibnamefont{Mukamel}},
  \bibinfo{journal}{Rev. Mod. Phys.} \textbf{\bibinfo{volume}{81}},
  \bibinfo{pages}{1665} (\bibinfo{year}{2009}).

\bibitem[{\citenamefont{{Campisi}
  et~al.}(2010{\natexlab{a}})\citenamefont{{Campisi}, {H\"anggi}, and
  {Talkner}}}]{Campisi11RMPXX}
\bibinfo{author}{\bibfnamefont{M.}~\bibnamefont{{Campisi}}},
  \bibinfo{author}{\bibfnamefont{P.}~\bibnamefont{{H\"anggi}}},
  \bibnamefont{and} \bibinfo{author}{\bibfnamefont{P.}~\bibnamefont{{Talkner}}}
  (\bibinfo{year}{2010}{\natexlab{a}}), \eprint{arXiv:1012.2268}.

\bibitem[{\citenamefont{{Campisi}
  et~al.}(2010{\natexlab{b}})\citenamefont{{Campisi}, {Talkner}, and
  {H{\"a}nggi}}}]{Campisi10PRL105}
\bibinfo{author}{\bibfnamefont{M.}~\bibnamefont{{Campisi}}},
  \bibinfo{author}{\bibfnamefont{P.}~\bibnamefont{{Talkner}}},
  \bibnamefont{and}
  \bibinfo{author}{\bibfnamefont{P.}~\bibnamefont{{H{\"a}nggi}}},
  \bibinfo{journal}{Phys. Rev. Lett.} \textbf{\bibinfo{volume}{105}},
  \bibinfo{pages}{140601} (\bibinfo{year}{2010}{\natexlab{b}}).

\bibitem[{\citenamefont{Morikuni and Tasaki}(2010)}]{Morikuni10arxiv}
\bibinfo{author}{\bibfnamefont{Y.}~\bibnamefont{Morikuni}} \bibnamefont{and}
  \bibinfo{author}{\bibfnamefont{H.}~\bibnamefont{Tasaki}}
  (\bibinfo{year}{2010}), \eprint{arXiv:1012.2753}.

\bibitem[{\citenamefont{Landau}(1932)}]{Landau32PZS2}
\bibinfo{author}{\bibfnamefont{L.~D.} \bibnamefont{Landau}},
  \bibinfo{journal}{Phys. Z. Sowjetunion} \textbf{\bibinfo{volume}{2}},
  \bibinfo{pages}{46} (\bibinfo{year}{1932}).

\bibitem[{\citenamefont{Zener}(1932)}]{Zener32PRSA137}
\bibinfo{author}{\bibfnamefont{C.}~\bibnamefont{Zener}},
  \bibinfo{journal}{Proc. R. Soc. A} \textbf{\bibinfo{volume}{137}},
  \bibinfo{pages}{696} (\bibinfo{year}{1932}).

\bibitem[{\citenamefont{St\"uckelberg}(1932)}]{Stueckelberg32HPA5}
\bibinfo{author}{\bibfnamefont{E.~C.~G.} \bibnamefont{St\"uckelberg}},
  \bibinfo{journal}{Helv. Phys. Acta} \textbf{\bibinfo{volume}{5}},
  \bibinfo{pages}{369} (\bibinfo{year}{1932}).

\bibitem[{\citenamefont{Majorana}(1932)}]{Majorana32NC9}
\bibinfo{author}{\bibfnamefont{E.}~\bibnamefont{Majorana}},
  \bibinfo{journal}{Nuovo Cimento} \textbf{\bibinfo{volume}{9}},
  \bibinfo{pages}{43} (\bibinfo{year}{1932}).

\bibitem[{\citenamefont{Holevo}(2001)}]{Holevo01Book}
\bibinfo{author}{\bibfnamefont{A.~S.} \bibnamefont{Holevo}},
  \emph{\bibinfo{title}{Statistical Structure of Quantum Theory}}
  (\bibinfo{publisher}{Springer}, \bibinfo{address}{Berlin},
  \bibinfo{year}{2001}).

\bibitem[{\citenamefont{Messiah}(1962)}]{Messiah62Book}
\bibinfo{author}{\bibfnamefont{A.}~\bibnamefont{Messiah}},
  \emph{\bibinfo{title}{Quantum Mechanics}} (\bibinfo{publisher}{North
  Holland}, \bibinfo{address}{Amsterdam}, \bibinfo{year}{1962}).

\bibitem[{\citenamefont{Andrieux and Gaspard}(2008)}]{Andrieux08PRL100}
\bibinfo{author}{\bibfnamefont{D.}~\bibnamefont{Andrieux}} \bibnamefont{and}
  \bibinfo{author}{\bibfnamefont{P.}~\bibnamefont{Gaspard}},
  \bibinfo{journal}{Phys. Rev. Lett.} \textbf{\bibinfo{volume}{100}},
  \bibinfo{pages}{230404} (\bibinfo{year}{2008}).

\bibitem[{\citenamefont{Vitanov}(1999)}]{Vitanov99PRA59}
\bibinfo{author}{\bibfnamefont{N.~V.} \bibnamefont{Vitanov}},
  \bibinfo{journal}{Phys. Rev. A} \textbf{\bibinfo{volume}{59}},
  \bibinfo{pages}{988} (\bibinfo{year}{1999}).

\bibitem[{\citenamefont{Misra and Sudarshan}(1977)}]{Misra77JMP18}
\bibinfo{author}{\bibfnamefont{B.}~\bibnamefont{Misra}} \bibnamefont{and}
  \bibinfo{author}{\bibfnamefont{E.~C.~G.} \bibnamefont{Sudarshan}},
  \bibinfo{journal}{J. Math. Phys.} \textbf{\bibinfo{volume}{18}},
  \bibinfo{pages}{756} (\bibinfo{year}{1977}).

\bibitem[{\citenamefont{Peres}(1980)}]{Peres80AJP48}
\bibinfo{author}{\bibfnamefont{A.}~\bibnamefont{Peres}}, \bibinfo{journal}{Am.
  J. Phys.} \textbf{\bibinfo{volume}{48}}, \bibinfo{pages}{931}
  (\bibinfo{year}{1980}).

\bibitem[{\citenamefont{Koshino and Shimizu}(2005)}]{Koshino05PREP412}
\bibinfo{author}{\bibfnamefont{K.}~\bibnamefont{Koshino}} \bibnamefont{and}
  \bibinfo{author}{\bibfnamefont{A.}~\bibnamefont{Shimizu}},
  \bibinfo{journal}{Phys. Rep.} \textbf{\bibinfo{volume}{412}},
  \bibinfo{pages}{191 } (\bibinfo{year}{2005}).

\bibitem[{\citenamefont{Facchi and Pascazio}(2008)}]{Facchi08JPA41}
\bibinfo{author}{\bibfnamefont{P.}~\bibnamefont{Facchi}} \bibnamefont{and}
  \bibinfo{author}{\bibfnamefont{S.}~\bibnamefont{Pascazio}},
  \bibinfo{journal}{J. Phys. A: Math. Theor.} \textbf{\bibinfo{volume}{41}},
  \bibinfo{pages}{493001} (\bibinfo{year}{2008}).

\bibitem[{\citenamefont{Peres}(1993)}]{Peres93Book}
\bibinfo{author}{\bibfnamefont{A.}~\bibnamefont{Peres}},
  \emph{\bibinfo{title}{Quantum Theory: Concepts and Methods}}
  (\bibinfo{publisher}{Kluwer}, \bibinfo{address}{Dordrecht},
  \bibinfo{year}{1993}).

\bibitem[{\citenamefont{Schmiedl et~al.}(2009)\citenamefont{Schmiedl,
  Dieterich, Dieterich, and Seifert}}]{Schmiedl09JSM09}
\bibinfo{author}{\bibfnamefont{T.}~\bibnamefont{Schmiedl}},
  \bibinfo{author}{\bibfnamefont{E.}~\bibnamefont{Dieterich}},
  \bibinfo{author}{\bibfnamefont{P.-S.} \bibnamefont{Dieterich}},
  \bibnamefont{and} \bibinfo{author}{\bibfnamefont{U.}~\bibnamefont{Seifert}},
  \bibinfo{journal}{J. Stat. Mech.: Theory Exp.}, \bibinfo{pages}{P07013}
  (\bibinfo{year}{2009}).

\bibitem[{\citenamefont{Sagawa and Ueda}(2010)}]{Sagawa10PRL104}
\bibinfo{author}{\bibfnamefont{T.}~\bibnamefont{Sagawa}} \bibnamefont{and}
  \bibinfo{author}{\bibfnamefont{M.}~\bibnamefont{Ueda}},
  \bibinfo{journal}{Phys. Rev. Lett.} \textbf{\bibinfo{volume}{104}},
  \bibinfo{pages}{090602} (\bibinfo{year}{2010}).

\bibitem[{\citenamefont{Zueco et~al.}(2008)\citenamefont{Zueco, H\"anggi, and
  Kohler}}]{ZuecoNJP08}
\bibinfo{author}{\bibfnamefont{D.}~\bibnamefont{Zueco}},
  \bibinfo{author}{\bibfnamefont{P.}~\bibnamefont{H\"anggi}}, \bibnamefont{and}
  \bibinfo{author}{\bibfnamefont{S.}~\bibnamefont{Kohler}},
  \bibinfo{journal}{New J. Phys.} \textbf{\bibinfo{volume}{10}},
  \bibinfo{pages}{115012} (\bibinfo{year}{2008}).

\bibitem[{\citenamefont{\'Alvarez et~al.}(2010)\citenamefont{\'Alvarez, Rao,
  Frydman, and Kurizki}}]{Alvarez10PRL105}
\bibinfo{author}{\bibfnamefont{G.~A.} \bibnamefont{\'Alvarez}},
  \bibinfo{author}{\bibfnamefont{D.~D.~B.} \bibnamefont{Rao}},
  \bibinfo{author}{\bibfnamefont{L.}~\bibnamefont{Frydman}}, \bibnamefont{and}
  \bibinfo{author}{\bibfnamefont{G.}~\bibnamefont{Kurizki}},
  \bibinfo{journal}{Phys. Rev. Lett.} \textbf{\bibinfo{volume}{105}},
  \bibinfo{pages}{160401} (\bibinfo{year}{2010}).

\end{thebibliography}

\end{document}